# Theoretical Investigation of (Zn, Co) co-Doped BaTiO$_3$ for Advanced Energy and Photonic Applications

*Zheng Kang[a], Mei Wu[a], Yiyu Feng[a], Jiahao Li[a], Jieming Zhang[a], Haiyi Tian[a], Ancheng Wang[b], Yunkai Wu*[a], Xu Wang*[c]*

a: College of Big Data and Information Engineering, Guizhou University, Guiyang, 550025, Guizhou, China

b: Department of Mathematics, The Chinese University of Hong Kong, Hong Kong SAR 999077, China

c: Key Laboratory of Advanced Manufacturing Technology, Ministry of Education, Guiyang, 550025, Guizhou, China

*Corresponding author. E-mail: ykwu@gzu.edu.cn

**Corresponding author. E-mail: xuwang@gzu.edu.cn

**Abstract:** In light of recent advancements in energy technology, there is an urgent need for lead-free barium titanate (BTO) -based materials that exhibit remarkable ferroelectric and photoelectric properties. Notwithstanding the considerable experimental advances, a theoretical understanding from the electron and atomic perspectives remains elusive. This study employs the generalized gradient approximation plane wave pseudopotential technique to investigate the structural, electronic, ferroelectric, and optical properties of (Zn, Co) co-doped BaTiO$_3$ (BZCT) based on density functional theory. The objective is to ascertain the extent of performance enhancement and the underlying mechanism of (Zn, Co) co-doping on barium titanate. Our findings reveal that the incorporation of (Zn, Co) into the BaTiO$_3$ lattice significantly augments the tetragonality of the unit cell. Moreover, the ferroelectric properties are enhanced, with a spontaneous polarization that is stronger than that observed in pure BTO, exhibiting excellent ferroelectricity. The results of the Hubbard+U algorithm indicate that the band gap of BZCT is reduced. Concurrently, the enhanced ferroelectric polarization increases the built-in electric field of the material, facilitating the separation of photogenerated carriers and improving optical absorption. Consequently, the optical absorption ability and photorefractive ability are effectively enhanced. BZCT, with its high spontaneous polarization and outstanding optical properties, can serve as a promising candidate material in the fields of energy storage and photovoltaics.

# I Introduction.

Barium titanate $BaTiO_3$(BTO), a member of the $ABO_3$ perovskite oxide family, is a prototypical lead-free ferroelectric material[1]. The subject of recent research has attracted a growing interest. Since its discovery in 1941, BTO has been the subject of considerable research interest[2]. Due to its numerous promising physical properties, including a high dielectric constant, positive resistivity temperature coefficient, high voltage tunability, piezoelectricity, ferroelectricity, low leakage current, and low dielectric dispersion, BTO has emerged as a versatile material for various applications within the electronics industr[3-10].

However, it also presents certain practical limitations due to its relatively low Curie temperature, which is around 120°C, a narrow range of tetragonal phase stability, a broad energy band gap, and a high dielectric constant at the Curie point. These characteristics can restrict its application in certain high-temperature or high-frequency electronic devices[11]. Significant research has been conducted to enhance the ferroelectric and optical characteristics of barium titanate, with the objective of expanding its potential applications. The prevailing research approach currently entails the synthesis of novel systems through the replacement of $Ba^{2+}$ or $Ti^{4+}$ with analogous ions of comparable dimensions, a process known as doping modification. A multitude of metal oxides have been employed in this manner to enhance the electronic and optical characteristics of the raw materials and expand their applications in optoelectronics[12, 13]. Given the evident significance of $BaTiO_3$-based materials, a substantial body of experimental and theoretical research has been conducted to investigate the modulation of their physicochemical and electrical properties through doping. Lu Wang et al.[14] disrupted the long-range ferroelectric ordering by incorporating $Li^+$ and $.Bi^{3+}$ into $BaTiO_3$ ceramics via solid-phase sintering. The resulting nanoclusters effectively suppressed the polarization and sustained the electrical strain, attaining the highest electrostriction coefficient of $0.0712 m^4/c^2$ to date among all known electrostrictive materials. Lois et al.[15] discovered that the Zn-doped $BaTiO_3$ system not only exhibits a linear decrease in lattice constant with respect to the Zn content but also demonstrated that this doped system is capable of providing enhanced ferroelectric and dielectric properties compared to the pure BTO. Additionally, there was a notable reduction in the bandgap. Anju et al[16]. formed a solid solution of $Sm_xBa_{1-x}TiO_3$ by doping $BaTiO_3$ with $Sm^{3+}$. The substitution of $Sm^{3+}$ results in lattice distortion of the grains due to the difference in the size of the substitutional ions, which enhances the tetragonal nature of the grains. $Sm^{3+}$ doping reduces dielectric loss and increases the dielectric constant, thereby enhancing the dielectric properties of the system. Additionally, $Sm^{3+}$ doping contributes to an increase in the carrier concentration and the formation of defects and vacancies in the material, which in turn leads to an enhancement in the spontaneous polarization of the system. The properties of barium zirconate titanate (BZT) ceramics are significantly influenced by varying amounts of zirconium substitution, resulting in the emergence of desirable piezoelectric, ferroelectric, and other electro-mechanical properties[17, 18]. The substitution of $Ca^{2+}$ in the A-site and $Zr^{4+}$ in the B-site of $ABO_3$ perovskites results in the formation of (Ba, Ca)(Zr, Ti)$O_3$, which alters the lattice parameters and causes a shift in the phase transition temperature and a broadening of the peak at the maximum value of the dielectric constant[19].

For the growing new energy industry, based on the high dielectric constant and the large spontaneous polarization of barium titanate, these barium titanate compounds have been greatly emphasized in a variety of applications, such as in photoelectrochemical systems used to increase the separation of carriers[20], in energy storage capacitors[21] or in the electronic ceramics industry[21-23]. However, in light of ongoing technological advancements, there is a growing need for ferroelectric materials that exhibit enhanced ferroelectric and optoelectronic properties. BaZnTiO$_3$ has been demonstrated to exhibit enhanced ferroelectric and dielectric properties relative to BTO, with minimal impact on the lattice constants[15]. Conversely, Co$^{4+}$ has been shown to markedly enhance the polarization properties of BTO[24].

Previously, the effects of co-doping BTO with Zn$^{2+}$ and Co$^{4+}$ ions on its ferroelectric and photovoltaic properties have not been extensively investigated. This is partly due to the limitations imposed by experimental conditions, which have hindered a detailed exploration of the material's electronic and band structures. To address this knowledge gap, we have utilized first-principles calculations to introduce Co$^{2+}$ and Zn$^{2+}$ ions into BTO crystals and assess their influence on the material's properties. This study is, to our knowledge, one of the first to systematically investigate the impact of (Zn, Co) co-doping on BTO ferroelectricity and the associated mechanisms, focusing on the local interactions, structural modifications, and the resulting electrical and optical properties.

## II. Calculation details

The doping system has been investigated using first-principles calculations and the supercell method. The density functional theory calculations are based on the Vienna ab initio simulation package (VASP)[25, 26]. The exchange-correlation energy of electrons was calculated under the generalized gradient approximation (GGA) using the Per - dew-Burke-Ernzerhof (PBE)method[27]. Select a 2 × 2 × 2 supercell containing 40 atoms, belonging to the P4mm space group, as shown in Fig. 1(a)[28]. And the initial lattice constants a=b= 3.99 Å and c= 4.01 Å. Based on the supercell, we introduced a Zn atom and a Co atom to replace the Ba and Ti atoms in the BTO supercell, respectively. As shown in Fig. 1(b). It is well known that DFT has problems in correctly describing the strong correlations between the d electrons, so the DFT + U method was used, and the 3d orbitals of the Ti atom and the 3d orbitals of the Co atom were corrected using the GGA + U method based on the method proposed in the literature[29] with the correction values of U = 9.4 eV and U = 5 eV, respectively[30, 31]. The cutoff value was chosen to be 500 eV. Using the Monkhorst-Pack method[32], a 5 × 5 × 5 grid of K-points centered on the gamma point was chosen for structure optimization and property calculations[33]. The convergence criterion for the interatomic interaction force is 2 × $10^{-2}$ eV/Å and for the system, energy is 1 × $10^{-5}$ eV/Å. The spontaneous polarization is calculated using the standard Berry-phase method.

## III, Results and Discussion

### 3.1. Geometry optimization.

Fig. 1 depicts the lattice models of pure BTO and BZCT, wherein elemental substitution was conducted with a single Zn atom and Co atom at the Ba site and Ti site, respectively. These calculations were performed using 40 atoms. The optimized lattice parameters of pure BTO and BZCT are presented in Table 1. The impact of (Zn, Co) on the structural properties of BTO materials is evaluated by examining the lattice parameters, cell angle, and tetragonality factor (c/a). The calculations yielded the following values for the lattice parameters of pure BTO: a = b = 3.96 Å and c = 4.04 Å. These values are in good agreement with those reported in previous experimental studies[34] and theoretical works[35]. The differences between our calculated lattice parameters and the previously reported theoretical and experimental lattice parameters are 0.025 Å and 0.000 Å, respectively, with an error of less than 3%. This indicates that our present work is reasonable.

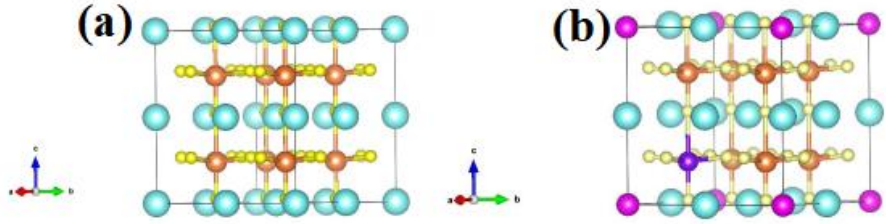

**Fig. 1** Supercell lattice model of (a)BTO and (b)BZCT (light blue for barium atoms, orange spheres for titanium atoms, purple spheres for cobalt atoms, magenta spheres for zinc atoms, yellow spheres for oxygen atoms)
.

**Table 1** Lattice parameters (Å), unit cell angle (in degrees), tetragonal factor (c/a), and volume (Å$^3$) for BTO and BZCT models

|  | a(Å) | b(Å) | C(Å) | α | β | γ | V(A$^3$) | c/a |
|---|---|---|---|---|---|---|---|---|
| BTO | 3.962 | 3.962 | 4.041 | 90° | 90° | 90° | 63.66 | 1.0172 |
| Theory | 4.025 | 4.025 | 4.066 | 90° | 90° | 90° | 65.39 | 1.009[35] |
| Experiment | 3.991 | 3.991 | 4.041 | 90° | 90° | 90° | 64.36 | 1.013[34] |
| BZTO | 3.94 | 3.94 | 4.14 | 90° | 90° | 90° | 64.26 | 1.05000 |
| BZCT | 3.944 | 3.944 | 4.052 | 90° | 89° | 90° | 62.87 | 1.02790 |

It is evident that doping results in a reduction in the lattice parameter of BZCT, an increase in the c lattice parameter, a slight rise in the c/a ratio, and an enhancement of the tetragonal nature of the lattice. Concurrently, the β angle, which is associated with the Zn atoms, undergoes a slight decline, and the lattice symmetry is diminished. The lattice aberration induced by (Zn, Co) doping is particularly pronounced. This is attributed to the difference in ion size between the dopant elements, Zn and Co, and the substituted ions. Additionally, the cell volume of the doped system is smaller than that of the pure BTO, indicating that the dopant elements induce lattice shrinkage. Following atomic relaxation, the optimized structure exhibits a smaller lattice volume, enhanced tetragonality, and relatively weakened symmetry, which is a contributing factor to the enhanced intrinsic polarization observed in BZCT ferroelectrics.

### 3.2. Electronic properties

The electronic properties of a material, such as its energy band structure and density of states(DOS), can reflect a multitude of physical properties and bonding characteristics. The study of the electronic energy band structure provides useful information for realizing paraelectricity and ferroelectricity, including semiconductor behavior.

In order to gain a deeper understanding of the underlying mechanism of lattice distortion, we selected the 100-plane and the 110-plane, which contain dopant elements in the form of Zn and Co atoms, for a charge density analysis. This is illustrated in Fig. 2, where the upward direction corresponds to the c-axis. As illustrated in Fig. 2, the Ti and Co ions in BZCT exhibit relaxation along the c-axis direction, deviating from the Ti-O facets in comparison to the pure BTO structure. Additionally, as illustrated in Fig. 2(c), there is an overlapping region of electron densities between the Zn ions and the adjacent O ions, indicating a robust interaction between Zn and O. This observation implies the presence of covalent bonding character in the Zn-O linkage. These results in the adjacent O2 atoms of the Zn ions undergoing reverse relaxation along the C-axis, which causes a significant zigzagging of the Ti-O plane in this layer of the O2 ions and an increase in the distortion of the oxygen octahedron. Consequently, this causes an increase in the displacement of Ti atoms from the center of the oxygen octahedron, which accounts for the observed enhancement in the intrinsic polarization of the doped system.

Furthermore, Fig. 2(b) illustrates that the electronic charge densities of Ti ions and Co ions overlap with the oxygen ions, forming covalent bonds. This phenomenon is analogous to the bonding observed between Zn ions and corresponding O2 ions, as depicted in Fig. 2(c). Furthermore, the nature of the bonding between the Ti ions and the O2 ions is markedly different, with no overlap of electronic layers between Ti-O2. It can thus be postulated that the promotion of oxygen octahedral distortion by the Ti-O planar fluctuation of BZCT, caused by Co-O chemical bonding and the formation of covalent bonds between Zn-O, represents the primary factors responsible for the enhancement of ferroelectricity and the improvement of other properties of BZCT.

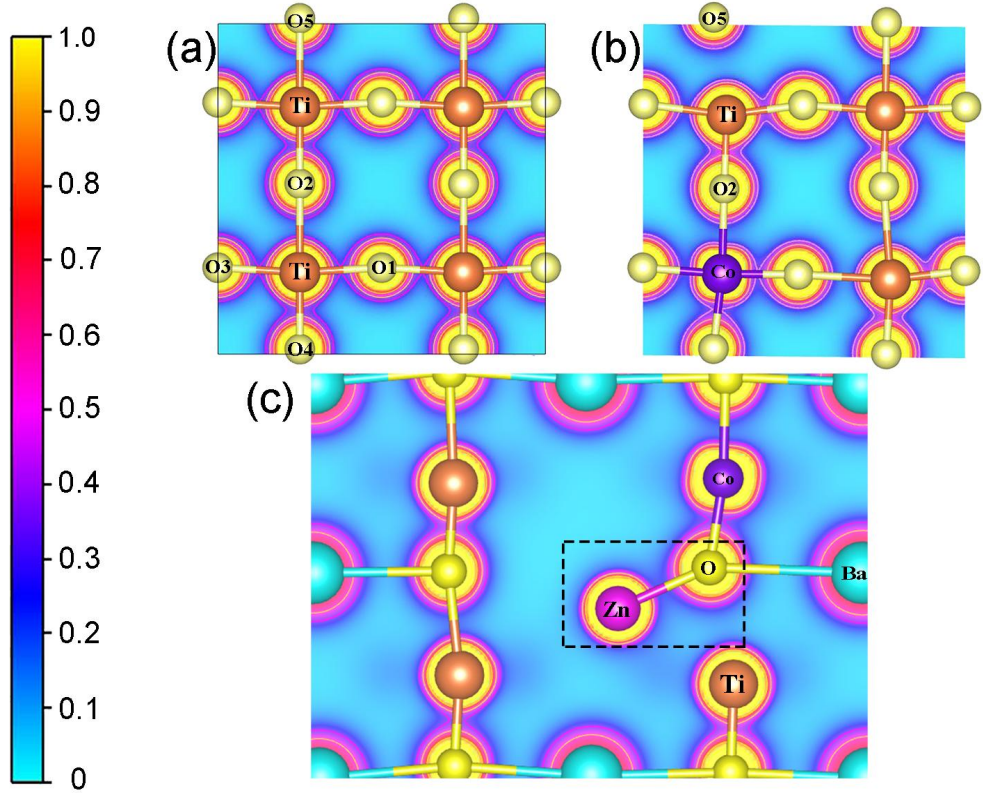

**Fig. 2** (100)plane charge density plots of (a)BTO,(b)BZCT, and (c) (110)plane charge density plots of BZCT.

To study the electronic structure of BTO perovskites and their doping modification changes, the electronic energy band structures of pure BTO and BZCT in the Brillouin zone along the high symmetry direction were calculated. The Fermi energy level was set to zero, which is shown as a gray dashed line in the figure. From Fig. 3(a), it can be seen that the CBM is located at the G-point due to the dominance of the Ti-3d state, while the VBM is located at the G-point at FL(0 eV), which is guided by the O-2p-state. The valence band tops (VBM) and conduction band bottoms (CBM) of the BTOs are located at highly symmetric G-points, indicating that the BTOs are direct bandgap semiconductors with an energy band value of 1.7918 eV, which is in close agreement with the reported values of 1.723 eV[36] and 1.778 eV[37]. Fig. 3(b) shows the electrified energy band structure of BZCT, where it is observed that the valence band top (VBM), predominantly constituted by the Ti-3d state, is situated at the X point, whereas the conduction band bottom (CBM), primarily comprising the O-2p state, is located at the Y point. This suggests that

BZCT is an indirect bandgap semiconductor. The bandgap of the BZCT calculated by GGA-PBE is 1.2847 eV.

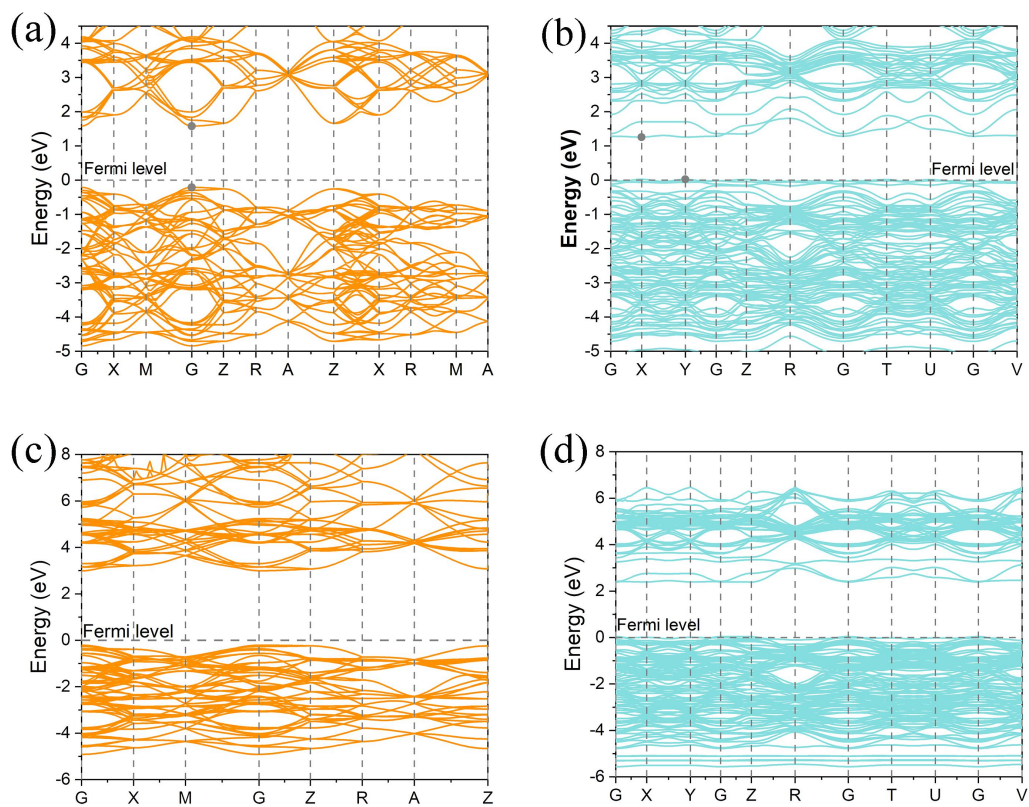

**Fig. 3** Band structure of (a)BTO，(b)BZCT without Hubbard U and band structure of (c)BTO，(d)BZCT with Hubbard U

The previous experimental study indicated that the band gap of BTO is approximately 3.2 eV[38]. This discrepancy can be attributed to the fact that the generalized gradient approximation (GGA) methodology employed in the calculations of p-d repulsion for cations and anions, as well as the estimation of band gaps, often results in an underestimation of the latter[39]. The observed trend of decreasing band gaps for BTO is consistent with the typical underestimation of density functional theory observed for another perovskite material, $SrTiO_3$[40]. To rectify this discrepancy, we employed the Hubbard+U algorithm to compute the revised electronic energy band structure of pure BTO and BZCT. The outcomes of this calculation are illustrated in Fig. 3(c) and Fig. 3(d). Following the correction by the Hubbard+U algorithm, the electronic energy band gap of pure BTO is 3.21 eV, which is in agreement with other experimental findings [38],[41]. The U-added algorithm of BZCT demonstrates that its energy bandwidth bandgap is 2.34 eV, a value that is smaller than that of pure BTO. The electrons and holes can be excited by lower electron energy, and due to its status as a direct bandgap semiconductor, the material is conducive to carrier migration. Consequently, the BZCT can be applied to optoelectronic materials. As this paper is concerned with the comparative alterations in properties prior to and following doping, along with the underlying mechanisms, the data will be used without the inclusion of U in the subsequent investigation.

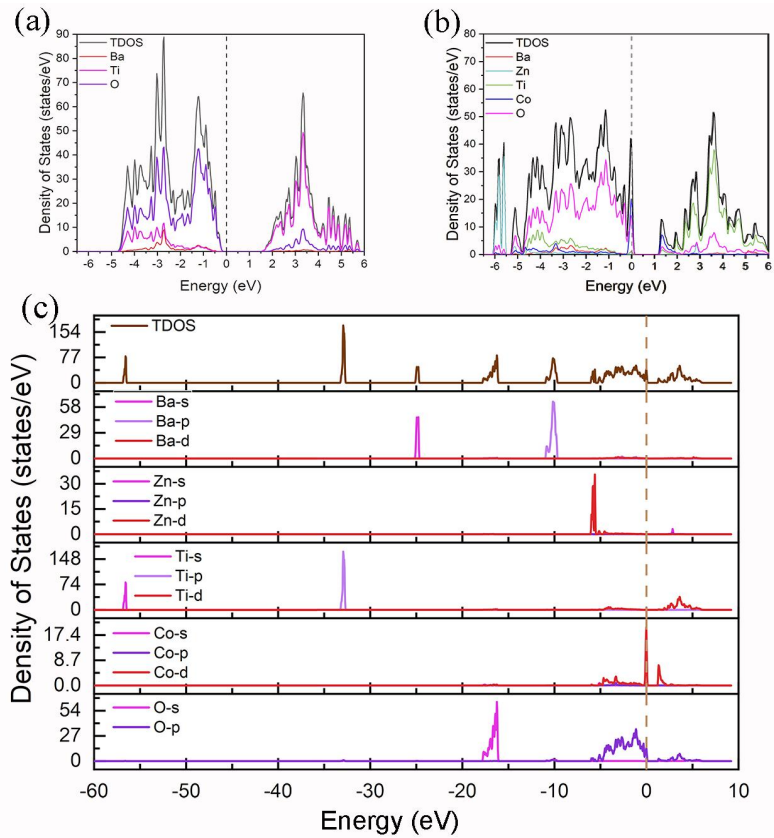

**Fig. 4** The total and partial density of states of (a) BTO-TDOS, (b) BZCT-TDOS, (c) BZCT-PDOS.

The total density of states (TDOS) and partial wave density of states (PDOS) for BTO and BCTO were calculated and are presented in Fig. 4. In the case of pure BTO, the energy range for the density of states (DOS) was selected to be between -6 eV and 6 eV. In the valence band (VB), the O-2p state is the primary contributor. In the conduction band (CB) region, the primary contributions are made by the Ti-3d and O-2p states. The hybridization of Ti-3d and O-2p states in the valence band (VB) and conduction band (CB) regions is a key factor contributing to the ferroelectricity observed in pure BTO[42]. Furthermore, the introduction of Zn and Co elements into the BTO cell results in the emergence of a new peak situated in close proximity to the Fermi energy level, specifically at 0 eV. The DOS plots of Fig. 4(b)(c) demonstrate that the primary contributions to this new state are Co-3d and O-2p states. The emergence of this peak results in the Fermi energy level of BCZT being situated in close proximity to the valence band. Concurrently, Co introduces a new electronic state at the base of the conduction band of BTO, thereby reducing the band gap of the energy band of BZCT.

The main contributors to the valence band and the conduction band of BCTO remain the Ti-3d and O-2p states, respectively. The impact of Zn doping is primarily manifested in the peak at -6eV in the Zn-3d valence band. In conclusion, the introduction of (Zn,Co) co-doping results in the emergence of new impurity energy levels within the material, leading to a downward shift in the conduction band and a shift in the Fermi energy levels towards the valence band. This phenomenon contributes to a reduction in the energy band gap of the entire system. Furthermore, the strong hybridization between Ti-3d and O-2p orbitals, as well as between Co-4d and O-2p orbitals, suggests that this is the factor responsible for the enhancement of ferroelectricity in the BZCT materials[42].

### 3.3 Electrical properties

Fig. 5 shows the energy difference and polarization function of pure BTO and BSZT, respectively. The energy polarization curve isitted by the phenomenological Landau–Ginzburg–Devonshire theory, and the equation is as follows.

$$\Delta G = \frac{1}{2}\alpha P^2 + \frac{1}{4}\beta P^2 + \frac{1}{6}\gamma P^2 \qquad (1)$$

where ΔGis the energy difference between the ferroelectric phase and the paraelectric phase, and α, β, γ are coefficient constants. The potential curves for both BTO and BCZT are well-fitted by the image-only Landau equation, respectively. The two minima in the double-trap potential curves correspond to the two stable polarisation states, the P+ state and the P- state, and, for each of these minima, the depth of the trap with respect to P=0 corresponds to the effective barrier for the reversal of the polarization. The P=0 state represents the paraelectric phase, which has no Ti displacement. For both cases of pure BTO and BZCT, the P+ and P- states are identical. It is generally accepted that the magnitude of the double-trap depth is proportional to the magnitude of the ferroelectric polarisation[43]. Therefore, analyzing the variation of the ferroelectric double-well potential depth will help to predict the ferroelectric phase transition and the response to the external electric field. As can be seen from Table 2, the double-well potential depth of BTO is 0.397 meV. The double-potential depth of BZCT is 0.766 meV, which is larger than that of BTO, and we predict that it is more difficult for BZCT to achieve polarisation switching. This is due to the larger ionic shift and stronger hybridization of Co-3d with O-2p orbitals and Ti-3d with O-2p orbitals. The spontaneous polarisation intensity of BZCT is 32.41 $\mu C/cm^2$, which is higher than that of pure BTO, which is 26.86 $\mu C/cm^2$. The changes in the spontaneous polarisation and the depth of the traps suggest that the spontaneous ferroelectricity of BCZT is higher than that of pure BTO ferroelectric is enhanced.

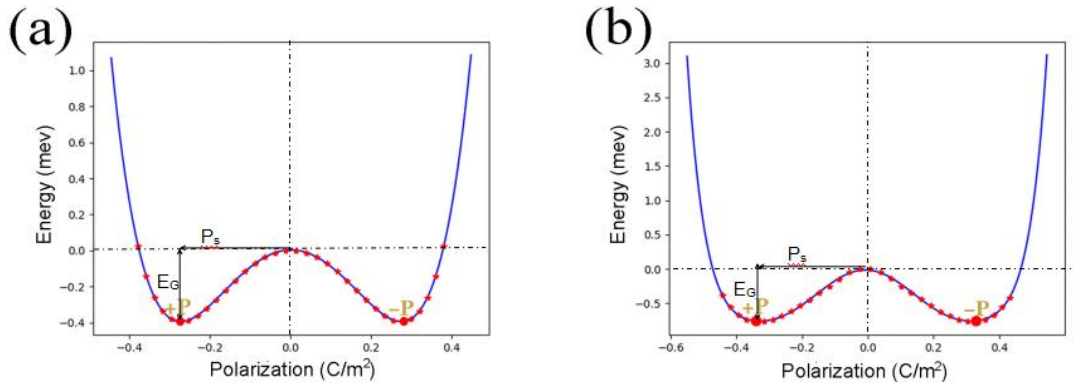

**Fig. 5** Double-well potentials of (a) pure BTO and (b) BZCT. The red dots are the total energy calculated by DFT and the blue line is from the phenomenological Landau model.

**Table 2** Polarization values $P_s$ and energy barrier $E_G$ of pure and doped $BaTiO_3$

|  | Polarization value $P_s$ ($\mu C/cm^2$) | Energy barrier $E_G$ (meV) |
|---|---|---|
| Pure BTO | 26.86 | 0.392 |
| BZCT | 32.41 | 0.766 |

In the realm of energy storage devices, superior performance is primarily determined by two critical metrics: enhanced energy storage density and optimal energy storage efficiency. The underpinnings of these attributes are typically reflected in the material's high spontaneous polarization and significant breakdown strength, according to the existing studies. Fig. 6 illustrates the transport properties of four distinct BTO systems, for which we have ascertained the conductivities. These calculations were grounded in the semi-empirical framework of the Boltzmann transport theory.

The conductivity σ can be obtained by integrating the distribution function(equation 1) over the whole space[44]:

$$\sigma_{\alpha\beta}(T;\mu) = \frac{1}{\Omega} \int \sigma_{\alpha\beta}(\varepsilon) [\frac{\partial f_{\mu}(T;\varepsilon)}{\partial \varepsilon}] \qquad (2)$$

where $f\mu(T,\varepsilon)$ is the Fermi-Dirac distribution function, μ is the chemical potential, and T is the temperature. When a constant relaxation time is known, it can be shown that the conductivity at RTA and RBA can be obtained for a certain chemical potential and temperature by calculating the energy band structure. Only their conductivities have been discussed qualitatively, so their relaxation time constants have not been estimated here.

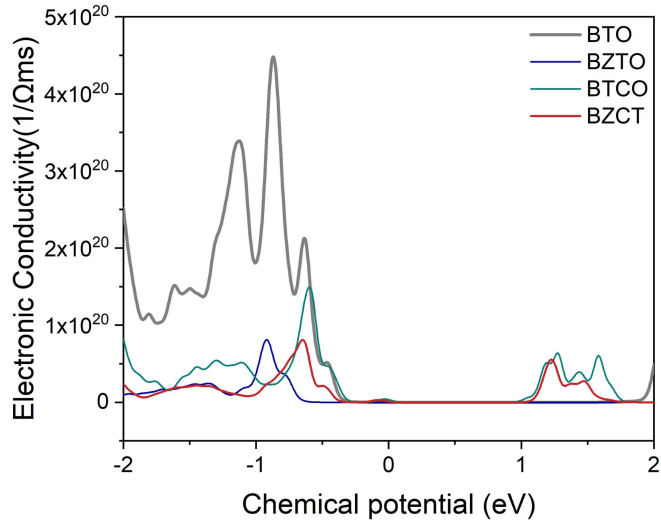

**Fig. 6** Electrical conductivity of BTO, BZTO, BTCO, and BZCT

From the figure, it can be seen that the transport properties behave differently between the pure BTO and doped BTO systems. A comparative analysis of the conductivity and relaxation time between the individual systems is illustrated in figure, where the horizontal coordinates are the maximum and minimum values of the chemical potential with respect to the Fermi energy level, We can see that the (Zn,Co) co-doped BZCT shows lower conductivity than the other three systems in a more stable state, and fewer carriers are transferred in the BZCT; and when the electrons in the system are in a higher energy state, i.e. the BZCT is in an excited state or there is an external energy input, such as sunlight irradiation, the conductivity is increased, which leads to more electrons participating in the conducting process, promoting carrier migration, which has good potential applications in the field of photovoltaics, When the electrons in the system are in a higher energy state, i.e., the BZCT is in an excited state or there is an external energy input, e.g., sunlight irradiation, the conductivity is increased, which leads to more electrons participating in the conducting process, which promotes carrier migration, which has good potential applications in the field of photovoltaics. This promotes carrier migration, which has good potential applications in the field of photovoltaics and is also consistent with the previous results of electronic energy band analysis. Conductivity describes the conductivity of the material, from the overall point of view, the conductivity of BZCT is weaker than the pure BTO system, indicating that the breakdown strength of BZCT is higher than that of pure BTO system, BZCT has a higher spontaneous polarization and higher breakdown strength than that of pure BTO[45], and BZCT is expected to be used in the field of energy storage.

### 3.4 Optical properties

In recent years, the effective utilization of solar energy has attracted the attention of many researchers [46, 47]. Ferroelectric photovoltaic materials have been demonstrated to exhibit an excellent photovoltaic effect, with photogenerated voltage not limited by the forbidden bandwidth (band gap) of the material itself. Furthermore, the photogenerated current can be regulated by the built-in electric field[48].

Accordingly, we have investigated the impact of modified ferroelectric characteristics on the optical attributes of the materials and explored the prospective applications of perovskite materials in photorefractive, optoelectronic, and solar cell (photovoltaic) domains.

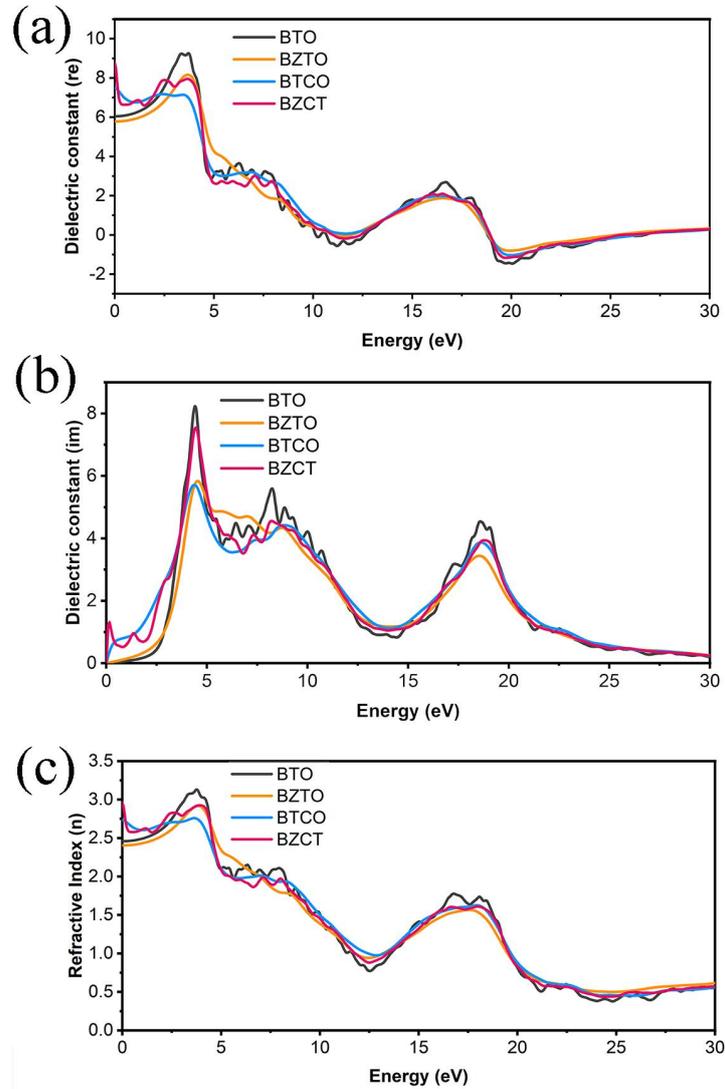

**Fig. 7** Optical properties of pure BTO and doped $BaTiO_3$ (a) real part of the complex dielectric function; (b) the imaginary part of the complex dielectric function; (c) refractive index.

In order to investigate the effect of (Zn, Co) co-doping on the optical properties of BTO, we calculated a series of optical properties of pure BTO, Zn-doped (BZTO), Co-doped (BTCO) and (Zn, Co) co-doped BTO(BZCT). These included the complex dielectric function, absorption function, loss function, reflectance, and refractive index, as illustrated in Fig. 7 and Fig. 8.

The complex dielectric function; $\varepsilon(\omega) = \varepsilon_1(\omega) + i\varepsilon_2(\omega)$ is divided into two parts: the real part of the complex dielectric function; denoted as $\varepsilon_1(\omega)$, and the imaginary part of the complex dielectric function, denoted as $\varepsilon_2(\omega)$.

The real part of the complex dielectric function, denoted as $\varepsilon_1(\omega)$, is indicative of the polarization properties of the material in question. As illustrated in Fig. 7(a), the real part of the complex dielectric function, denoted as $\varepsilon_1(\omega)$, is dependent on the incident photon energy. With an increase in the incident photon energy, the value of $\varepsilon_1(\omega)$ subsequently decreases, indicating a reduction in the material polarization property. As illustrated in Fig. 7(a), the polarization properties of BZCT (8.76) and BTCO (7.47) are superior to those of pure BTO(6.04) at 0.0 eV, suggesting that Co doping has a beneficial impact on the polarization properties. The pure and doped systems approach the minimum value of $\varepsilon_1(\omega)$ at a photon energy of approximately 19.0 eV, respectively, and exhibit a slight increase thereafter.

In contrast, the imaginary part of the complex dielectric function, represented as $\varepsilon_2(\omega)$, is associated with the energy dissipation observed within the system. As illustrated in Fig. 7(b), the imaginary component of the dielectric function, $\varepsilon_2(\omega)$, of the doped system exhibits an increase for all doped systems between 0 and 2.89 eV, indicating that the energy dissipation of the doped system is elevated. In particular, the imaginary part $\varepsilon_2(\omega)$ is higher for the (Zn, Co) and Co doping cases than for pure BTO and BZTO at lower incident photon energies, including the peak in the visible range of 1.65-3.10 eV. Furthermore, the energy dissipation is higher in the low-energy region for BZCT and BTCO. The pure BTO exhibits superior energy dissipation characteristics in the medium and high energy regions when compared to the doped system.

The refractive indices (n($\omega$)) of pure BTO and doped systems are demonstrated in Fig. 7(c),. The refractive indices of the refractive spectra of these materials in the infrared, visible, and most of the ultraviolet ranges are greater than 1. Co-doping has been observed to increase the static refractive indices of the pure BTOs, with the static refractive indices of the BZTOs, BTCOs, and BZCTs being 2. 40, 2.73, and 2.92, respectively. When the refractive indices exceed 1, photons encountering the material are decelerated due to electron interaction, resulting in a higher refractive index[49]. Materials with a refractive index of 1 or greater are considered transparent to incident light. Therefore, BZCT is transparent to incident light below 11.03 eV and opaque to incident light above this value. In general, any process that increases the electron density of a material will also result in an increase in the refractive index [50]. The effective enhancement of BTO refractive index properties by Co/Zn co-doping indicates that BZCT may be a suitable material for use in photorefractionation.

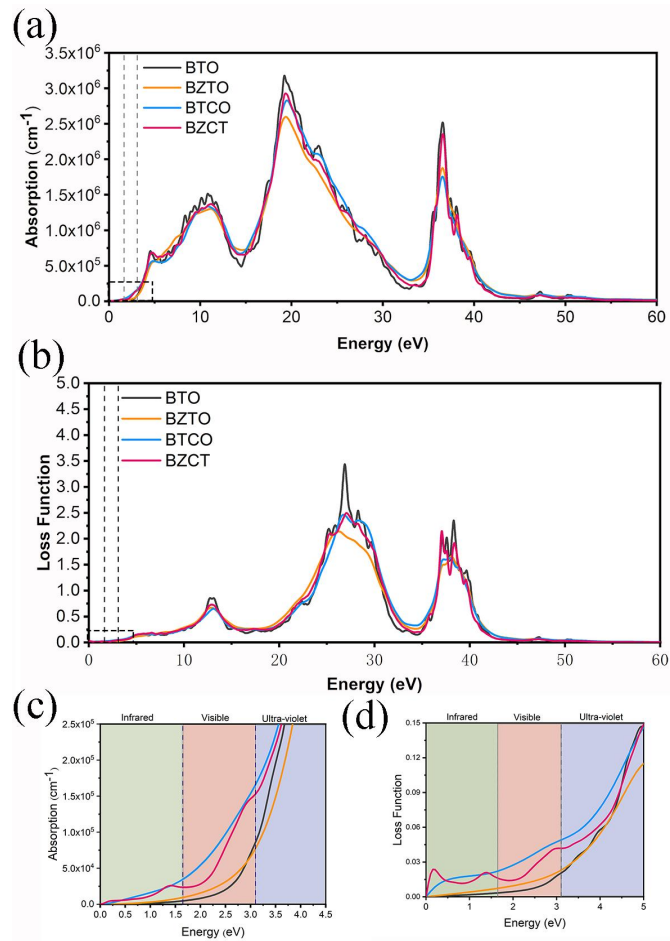

**Fig. 8** Optical properties of pure BTO and doped BaTiO$_3$ (a) absorption spectra; (b) loss function; (c) absorption spectrum in the region of 0-4.5 eV; (d) loss function in the region of 0-5 eV.

The absorption spectra α(ω) of BTO, BZTO, BTCO, and BZCT are illustrated in Fig. 8(a). At an incident photon energy of 19.21 eV, a pronounced absorption peak is observed for each system. In comparison to BTO, BZTO, BTCO, and BZCT, the latter exhibit lower absorption coefficients at the peak. The pure BTO system, in particular, demonstrates a pronounced absorption of electromagnetic radiation energy in the vicinity of 19 eV, which can be classified as occurring in the medium-energy region. In the photon energy range of 20 to 40 eV, the systems display comparable absorption spectra. In the low-energy range (0 – 10 eV), BZCT and BTCO display enhanced absorption characteristics. In comparison to the pure BTO system, the BZCT system demonstrates the capacity to markedly enhance the material's light absorption capability. The BZCT system exhibits superior light absorption capabilities in the visible and low UV regions, with a broader absorption spectral range than that of the pure BTO system. In the deep UV high-energy region of the absorption spectrum, the absorption peak of BZTO is stronger than that of BTCO. In conclusion, the introduction of a Co dopant in pure BTO shifts the absorption characteristics of the material to low frequency, while the addition of a Zn dopant increases the absorption characteristics of the material at high frequency. In light of the aforementioned findings, BZCT emerges as a promising candidate for utilization as an absorber layer in solar cells.

When electromagnetic radiation is incident on a material, a portion of the energy is lost within the material[51]. The energy loss is quantified by a loss function, as illustrated in Fig. 8(b). The highest peaks of the loss functions were observed in the pure system, and the energy loss peaks of the intrinsic and doped materials were obtained at the corresponding photon energies of 26-28 eV. It can be seen that the loss functions of the infrared region of BTCO and BZCT are also high. The energy loss of the doped system is less pronounced than that of the pure BTO system at all three peaks, indicating an effective enhancement of the optical properties by the doped system.

## III. CONCLUSION

The theoretical characterization of the crystal structure, electronic properties, ferroelectric, and optical properties of the (Zn, Co) co-doped BTO system was conducted through first-principle calculations, and the intrinsic mechanism of (Zn, Co) co-doping on the improvement of BTO properties was investigated. The results of the structure optimization demonstrate that the co-doping of (Zn,Co) results in a reduction of structural symmetry and an enhancement of lattice tetragonality in BaTiO$_3$. The band gap of the co-doped system of BZCT exhibits a notable reduction due to the addition of the impurities. The introduction of impurity energy levels in comparison to the pure BTO material results in a transition from a direct to an indirect band gap. The incorporation of Hubbard's energy corrects the band gap of the electronic energy bands from 3.20 eV to 2.34 eV.

The calculation of the density of states indicates the formation of strong force orbital hybridization between the Zn-3d and O-2p states, as well as between the Co-3d and O-2p states. It can be concluded that the Co-3d and O-2p states are the primary contributors to the shift in the conduction band level at the Fermi energy. In the charge density, covalent bonds are formed between Zn and O and Co and O, which represents a significant factor and direct manifestation of the lattice distortion. The deviation of the Ti-O planes results in a notable change in the oxygen octahedron, leading to a more pronounced spontaneous polarization of BZCT. This enhanced ferroelectricity makes BZCT a promising material for energy storage devices. It is noteworthy that the enhanced ferroelectricity improves the optical properties of BZCT, which exhibits enhanced light absorption in the visible and low ultraviolet regions. This excellent light absorption property makes BZCT a promising candidate in the field of ferroelectric photovoltaics and photocatalysis. The alterations in electrical and optical characteristics resulting from (Zn, Co) co-doping will expand the scope of applications for BaTiO$_3$ ferroelectric materials in the domains of energy storage and optics. Moreover, they will furnish a theoretical foundation for the investigation of the properties of perovskite materials and device fabrication in the future.

**Authorship contribution statement**

**Zheng Kang :** Writing - Original Draft, Conceptualization, Data curation, Visualization, Formal analysis. **Mei Wu:** Writing - Original Draft, Visualization, Software**. Yiyu Feng:** Original draft, Visualization. **Jiahao Li:** Software, Visualization.**Jieming Zhang:** Validation**,** Original draft. **Haiyi Tian**: Software,

Visualization. **Ancheng Wang:** Review, Editing. **Yunkai Wu:** Conceptualization, Review, Editing, and Funding acquisition. **Xu Wang:** Supervision, Review and Editing.

## Data availability

Data will be made available on request.

## Acknowledgments

The work was supported by the Young Scientists Fund of the National Natural Science Foundation of China (Grant No.62305075), and Science and Technology Plan Project of Guizhou Province (Grant No. QianKeHeJiChu-ZK[2024]YiBan032)

## IV. References

[1] Y. Ma, Y. Li, H. Wang, M. Wang, J.W.P. Envelope, High performance flexible photodetector based on 0D-2D perovskite heterostructure, Chip  (2022).
[2] C.A. Randall, R.E. Newnham, L.E. Cross, History of the First Ferroelectric Oxide, $BaTiO_3$, Web Source (2011).
[3] Scott, F. J., Data storage. Multiferroic memories, Nature Materials 6(4) (2007) 256.
[4] M.Q. Cai, Y.J. Zhang, Z. Yin, M.S. Zhang, First-principles study of structural and electronic properties of $BaTiO_3$ (001) oxygen-vacancy surfaces, Phys.rev.b 72(7) (2005) 075406(1-6).
[5] J.W. Lee, K. Eom, T.R. Paudel, B. Wang, H. Lu, H. Huyan, S. Lindemann, S. Ryu, H. Lee, T.H. Kim, In-plane quasi-single-domain $BaTiO_3$ via interfacial symmetry engineering, Nature Communications (2021).
[6] Y. Bai, X. Han, X.C. Zheng, L. Qiao, Both High Reliability and Giant Electrocaloric Strength in $BaTiO_3$ Ceramics, Scientific Reports 3 (2013) 2895.
[7] H. Lu, C.W. Bark, D.E.D.L. Ojos, J. Alcala, A. Gruverman, Mechanical Writing of Ferroelectric Polarization, Science 336(6077) (2012) 59-61.
[8] Hoffmann, Michael, J., Glaum, Julia, Genenko, Yuri, A., Albe, Karsten, Mechanisms of aging and fatigue in ferroelectrics, Materials Science & Engineering B Solid State Materials for Advanced Technology  (2015).
[9] H. Shen, K. Xia, P. Wang, R. Tan, The electronic, structural, ferroelectric and optical properties of strontium and zirconium co-doped $BaTiO_3$: First-principles calculations, Solid State Communications (2022).
[10] D.S. Fu, S. Hao, J.L. Li, L.S. Qiang, Effects of the penetration temperature on structure and electrical conductivity of samarium modified $BaTiO_3$ powders, Journal of Rare Earths 29(2) (2011) 164-167.
[11] B.C. Keswani, D. Saraf, S.I. Patil, A. Kshirsagar, A.R. James, Y.D. Kolekar, C.V. Ramana, Role of A-site Ca and B-site Zr substitution in $BaTiO_3$ lead-free compounds: Combined experimental and first principles density functional theoretical studies, J. Appl. Phys. 123(20) (2018) 16.
[12] M. Rizwan, A. Ayub, M. Shakil, Z. Usman, S. Gillani, H. Jin, C. Cao, Putting DFT to trial: For the exploration to correlate structural, electronic and optical properties of M-doped (M=Group I, II, III, XII, XVI) lead free high piezoelectric c-$BiAlO_3$, Materials Science and Engineering B-advanced Functional

Solid-state Materials 264 (2021) 114959.

[13] M. Benyoussef, H. Zaari, J. Belhadi, Y. El Amraoui, H. Ez-Zahraouy, A. Lahmar, M. El Marssi, Effect of rare earth on physical properties of $Na_{0.5}Bi_{0.5}TiO_3$ system: A density functional theory investigation, Journal of Rare Earths 40(3) (2022) 473-481.

[14] L. Wang, H. Qi, S.Q. Deng, L.Z. Cao, H. Liu, S.X. Hu, J. Chen, Design of superior electrostriction in $BaTiO_3$-based lead-free relaxors via the formation of polarization nanoclusters, InfoMat 5(1) (2023) 11.

[15] L.H.D.S. Lacerda, R.A.P. Ribeiro, A.M.D. Andrade, S.R.d. Lazaro, Zn-doped $BaTiO_3$ Materials: A DFT Investigation for Optoelectronic and Ferroelectric Properties Improvement, Revista Processos Químicos 9(18) (2015) 274-280.

[16] A. Kumari, K. Kumari, F. Ahmed, A. Alshoaibi, P.A. Alvi, S. Dalela, M.M. Ahmad, R.N. Aljawfi, P. Dua, A. Vij, S. Kumar, Influence of Sm doping on structural, ferroelectric, electrical, optical and magnetic properties of $BaTiO_3$, Vacuum 184 (2021) 14.

[17] C. Ciomaga, M. Viviani, M.T. Buscaglia, V. Buscaglia, L. Mitoseriu, A. Stancu, P. Nanni, Preparation and characterisation of the $Ba(Zr,Ti)O_3$ ceramics with relaxor properties, Journal of the European Ceramic Society 27(13-15) (2007) 4061-4064.

[18] V.V. Shvartsman, D.C. Lupascu, Lead-Free Relaxor Ferroelectrics, Journal of the American Ceramic Society 95(1) (2012) 1-26.

[19] S.M. Neirman, The Curie point temperature of $Ba(Ti_{1-x}Zr_x)O_3$ solid solutions, Journal of Materials Science 23(11) (1988) 3973-3980.

[20] Y. Liu, Z. Wang, C. Lin, J. Zhang, J. Feng, B. Hou, W. Yan, M. Li, Z. Ren, Spontaneous polarization of ferroelectric heterostructured nanorod arrays for high-performance photoelectrochemical cathodic protection, Applied Surface Science: A Journal Devoted to the Properties of Interfaces in Relation to the Synthesis and Behaviour of Materials  (2023).

[21] D. Li, X. Jiang, H. Hao, J. Wang, Q. Guo, L. Zhang, Z. Yao, M. Cao, H. Liu, Amorphous/Crystalline Engineering of $BaTiO_3$-Based Thin Films for Energy-Storage Capacitors, ACS Sustainable Chemistry & Engineering (10-4) (2022).

[22] Scott, F. J., Applications of Modern Ferroelectrics, Science  (2007).

[23] H. Xue, Y. Peng, Q. Jing, J. Zhou, G. Han, W. Fu, Sensing with extended gate negative capacitance ferroelectric field-effect transistors, Chip 3(1) (2024).

[24] L. Padilla-Campos, D.E. Diaz-Droguett, R. Lavín, S. Fuentes, Synthesis and structural analysis of Co-doped $BaTiO_3$, Journal of Molecular Structure 1099 (2015) 502-509.

[25] G.K. A, J.F. b, Efficiency of ab-initio total energy calculations for metals and semiconductors using a plane-wave basis set - ScienceDirect, Computational Materials Science 6(1) (1996) 15-50.

[26] G.G. Kresse, J.J. Furthmüller, Efficient Iterative Schemes for Ab Initio Total-Energy Calculations Using a Plane-Wave Basis Set, Physical review. B, Condensed matter 54 (1996) 11169.

[27] J.P. Perdew, K. Burke, M. Ernzerhof, Generalized Gradient Approximation Made Simple, Physical Review Letters 77(18) (1998) 3865-3868.

[28] K. Momma, F. Izumi, VESTA3 for three-dimensional visualization of crystal, volumetric and morphology data, Journal of Applied Crystallography 44 (2011) 1272-1276.

[29] W. Setyawan, S. Curtarolo, High-throughput electronic band structure calculations: challenges and tools, Computational Materials Science 49(2) (2010) 299-312.

[30] Lan, Guoqiang, Song, Jun, Yang, Zhi, A linear response approach to determine Hubbard U and its application to evaluate properties of $Y_2B_2O_7$, B = transition metals 3d, 4d and 5d, Journal of Alloys & Compounds An Interdisciplinary Journal of Materials Science & Solid State Chemistry & Physics

(2018).

[31] W. Setyawan, R.M. Gaume, S. Lam, R.S. Feigelson, S. Curtarolo, High-Throughput Combinatorial Database of Electronic Band Structures for Inorganic Scintillator Materials, ACS Comb. Sci. 13(4) (2011) 382-390.

[32] Chadi, J. D., Special points for Brillouin-zone integrations, Physical Review B 16(4) (1977) 1746-1747.

[33] V. Wang, N. Xu, J.C. Liu, G. Tang, W.T. Geng, VASPKIT: A user-friendly interface facilitating high-throughput computing and analysis using VASP code, Computer Physics Communications 267 (2021).

[34] N. Yasuda, H. Murayama, Y. Fukuyama, J. Kim, M. Takata, X-ray diffractometry for the structure determination of a submicrometre single powder grain, Journal of Synchrotron Radiation 16(Pt 3) (2009) 352-357.

[35] E. Ching-Prado, Stress dependence of structure, electronic and optical properties of $BaTiO_3$ from WC, VdW-DF-C09 and HSE functional calculations, Ferroelectrics 535(1) (2018) 171-182.

[36] M. Rizwan, Hajra, I. Zeba, M. Shakil, Z. Usman, Electronic, structural and optical properties of $BaTiO_3$ doped with lanthanum (La): Insight from DFT calculation, Optik - International Journal for Light and Electron Optics 211 (2020) 164611.

[37] M.H. Taib, M.K. Yaakob, N.H. Hussin, M.H. Samat, O.H. Hassan, M. Yahya, Structural, Electronic and Optical Properties of $BaTiO_3$ and $BaFeO_3$ From First Principles LDA+U Study, 2016.

[38] S. Dahbi, N. Tahiri, O. El Bounagui, H. Ez-Zahraouy, Electronic, optical, and thermoelectric properties of perovskite $BaTiO_3$ compound under the effect of compressive strain, Chem. Phys. 544 (2021) 6.

[39] M. Maraj, A. Fatima, S.S. Ali, U. Hira, M. Rizwan, Z. Usman, W.H. Sun, A. Shaukat, Taming the optical response via (Ca:Zr) co-doped impurity in c-$BaTiO_3$: A comprehensive computational insight, Mater. Sci. Semicond. Process 144 (2022) 9.

[40] U.S. Shenoy, H. Bantawal, D.K. Bhat, Band Engineering of $SrTiO_3$: Effect of Synthetic Technique and Site Occupancy of Doped Rhodium, Journal of Physical Chemistry C 122(48) (2018) 27567-27574.

[41] R. Khenata, M. Sahnoun, H. Baltache, M. Rerat, A.H. Rashek, N. Illes, B. Bouhafs, First-principle calculations of structural, electronic and optical properties of $BaTiO_3$ and $BaZrO_3$ under hydrostatic pressure, Solid State Communications 136(2) (2005) 120-125.

[42] Cohen, E. Ronald, Origin of ferroelectricity in perovskite oxides, Nature 358(6382) (1992) 136-138.

[43] H. De Raedt, K. Michielsen, Computational Methods for Simulating Quantum Computers, Physics (2004).

[44] L.Y. Hao, E.G. Fu, First-principles calculation on the electronic structures, phonon dynamics, and electrical conductivities of $Pb_{10}(PO_4)_6O$ and $Pb_9Cu(PO_4)_6O$ compounds, J. Mater. Sci. Technol. 173 (2024) 218-224.

[45] W.B. Gao, M.W. Yao, X. Yao, Achieving Ultrahigh Breakdown Strength and Energy Storage Performance through Periodic Interface Modification in $SrTiO_3$ Thin Film, ACS Appl. Mater. Interfaces 10(34) (2018) 28745-28753.

[46] M. Ahmadi, L. Collins, A. Puretzky, J. Zhang, J.K. Keum, W. Lu, I. Ivanov, S.V. Kalinin, B. Hu, Exploring Anomalous Polarization Dynamics in Organometallic Halide Perovskites, Advanced Materials 30(11) (2018) 1705298.1-1705298.10.

[47] K. Wojciechowski, S.D. Stranks, A. Abate, G. Sadoughi, H.J. Snaith, Heterojunction Modification for Highly Efficient Organic-Inorganic Perovskite Solar Cells, ACS Nano 8(12) (2014).

[48] K.X. Guo, R.F. Zhang, Z. Fu, L.Y. Zhang, X. Wang, C.Y. Deng, Regulation of Photovoltaic Response in ZSO-Based Multiferroic BFCO/BFCNT Heterojunction Photoelectrodes via Magnetization and Polarization ACS Appl. Mater. Interfaces 13(37) (2021) 45118-45118.

[49] C.B. Samantaray, H. Sim, H. Hwang, The electronic structures and optical properties of $BaTiO_3$ and $SrTiO_3$ using first-principles calculations, Microelectronics Journal 36(8) (2005) 725-728.

[50] V.B. Parmar, D. Raval, S.K. Gupta, P.N. Gajjar, A.M. Vora, $BaTiO_3$ perovskite for optoelectronics application: A DFT study, Materials Today: Proceedings (2023).

[51] S.S.A. Gillani, R. Ahmad, d. Islah u, M. Rizwan, M. Shakil, M. Rafique, G. Murtaza, H.B. Jin, First-principles investigation of structural, electronic, optical and thermal properties of Zinc doped $SrTiO_3$, Optik 201 (2020) 9.